# SocialHBC: Social Networking and Secure Authentication using Interference-Robust Human Body Communication

Shreyas Sen
School of Electrical and Computer Engineering (ECE), Purdue University
shreyas@purdue.edu

## ABSTRACT

With the advent of cheap computing through five decades of continued miniaturization following Moore's Law, wearable devices are becoming increasingly popular. These wearable devices are typically interconnected using wireless body area network (WBAN). Human body communication (HBC) provides an alternate energy-efficient communication technique between on-body wearable devices by using the human body as a conducting medium. This allows order of magnitude lower communication power, compared to WBAN, due to lower loss and broadband signaling. Moreover, HBC is significantly more secure than WBAN, as the information is contained within the human body and cannot be snooped on unless the person is physically touched. In this paper, we highlight applications of HBC as (1) Social Networking (e.g. LinkedIn/Facebook friend request sent during Handshaking in a meeting/party), (2) Secure Authentication using human-human or human-machine dynamic HBC and (3) ultra-low power, secure BAN using intra-human HBC. One of the biggest technical bottlenecks of HBC has been the interference (e.g. FM) picked up by the human body acting like an antenna. In this work, for the first time, we introduce an integrating dual data rate (DDR) receiver technique, that allows notch filtering (>20 dB) of the interference for interference-robust HBC.

## CCS Concepts

• **Hardware** → **Communication hardware, interfaces and storage** • **Security and privacy** → **Human and societal aspects of security and privacy** • **Human-centered computing** → **Human computer interaction (HCI).**

## Keywords

Human Body Communication (HBC), Social Networking, Body Coupled Communication (BCC), Secure Authentication, Ultra-Low Power (ULP), Interference tolerance, Resettable Integrator, Adaptive Notch Filter, Integrating DDR Receiver

## 1. INTRODUCTION

The continuous reduction of size of unit computing [1], has propelled the growth of *wearable* sensors and computing devices (e.g. Fitness trackers, Smart watches). This increasing growth of the wearable market is expected to grow to 600 million by 2020 [2]. Soon, Human Body will become a platform for *interconnected* wearable smart devices, which will aid and improve human quality of life. This calls for efficient ways to *connect* these wearable devices on the human body. Moreover, since each individual will be "wearing" a huge amount of information on their body (Human Intranet), they can now transmit this information to other humans or machines (Human Internet) at their will or use this information for secure authentication. Such on-body wearable devices are typically interconnected using WBAN. Human Body Communication (HBC) has recently emerged as a strong contender for this human body network, as it provides ultra-low power (ULP) and increased security, compared to WBAN. ULP is achieved as human body is used as a conducting medium, which exhibits significantly lower loss than radio frequency propagation through air. HBC is more secure as the information is contained within the human body and cannot be snooped on unless the person is physically touched, unlike WBAN, where the wireless signals can be easily snooped on by an attacker.

HBC was first introduced in the pioneering work [3] from MIT. The authors proposed capacitive near-field coupling and human body coupled conduction. The earth's ground was used as the reference (return path). This work provided a simplified electrical model of the HBC network, which treated the human body as a single node i.e. a perfect conductor. Since then, other methods of HBC, such as Galvanic coupling [4] has been explored. This requires direct skin contact and hence is less widely used compared to the capacitive coupled HBC. The authors in [5] used capacitive coupled HBC, along with an electro-optic sensor to increase sensitivity. In [6], authors from Philips Research provides a detailed overview of the progress of HBC up to 2008. Significant progress [7] has been made on modeling the HBC channel, such that it matches the measured characteristics over a wide frequencies and distances.

The human body acts as an antenna [8] at the FM frequency band. This has been the biggest bottleneck in high-speed ULP HBC implementation. Signaling techniques that allow to circumvent the interference, such as adaptive frequency hopping (AFH) [9] and fixed narrowband signaling [10] have been proposed. However, till date there has been no way to suppress the interference other than avoiding it using adaptive/fixed narrowband signaling, which leads to energy-inefficient implementation and needs bulky filters. In this work, we propose an adaptive broadband NRZ signaling scheme, which suppresses the undesired interference by using resettable integration with dual data rate (DDR) NRZ receiver. The theory supporting this technique is developed along with results to show the efficacy of the proposed technique under signal to interference ratio as high as -23 dB, i.e. successful HBC transmission is achieved even when the interference strength is 23 dB higher than the received signal, without use of any bulky filters.

The rest of the paper is organized as follows. Section 2 presents the application of HBC. Section 3 provides the details of interference-robust HBC. Results are presented in Section 4, with conclusions in Section 5.

## 2. APPLICATIONS OF HBC

The following subsections will highlight some of the most promising applications of HBC, such as social networking, secure authentication and ULP, secure BAN.

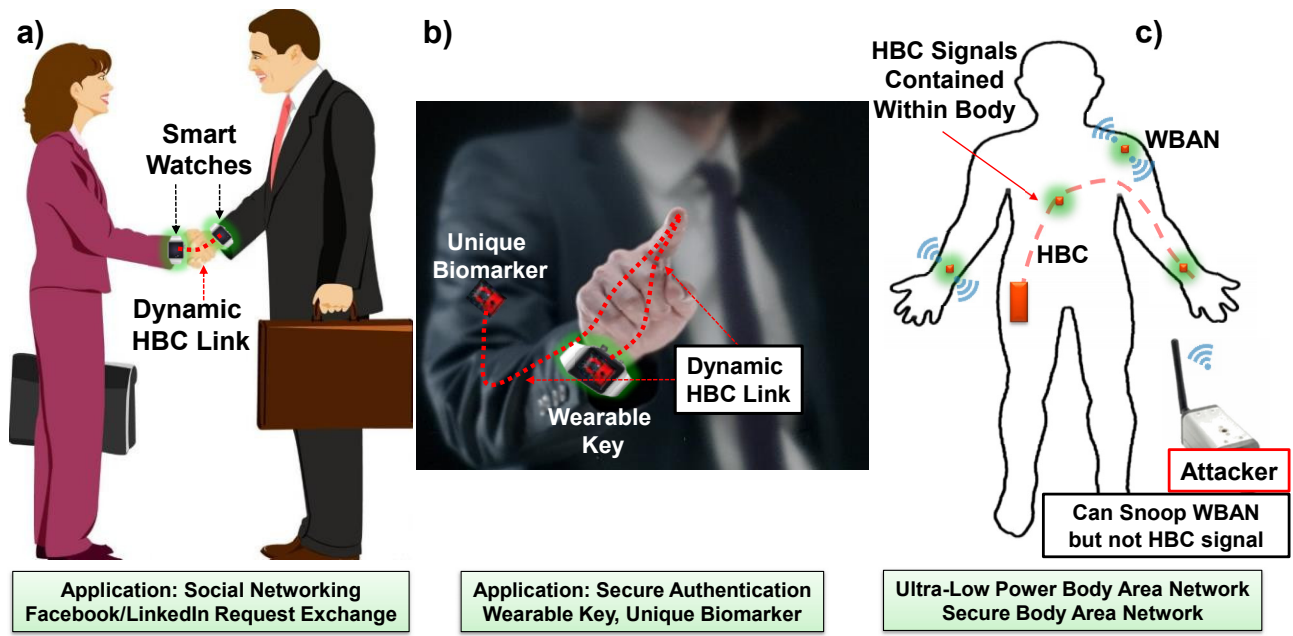

Figure 1: Applications of Human Body Communication (HBC). Dynamic HBC a) Social Networking: Facebook/LinkedIn request exchanged through the human body between smartwatches, during Handshaking in a Meeting/Party b) Secure Authentication: Secure wearable keys on smartwatch or other wearable devices could be used in addition to fingerprint for additional layer of security using the human body to communicate it during touch based authentication c) Static HBC allows ultra-low power body area network (BAN) due to low loss, broadband HBC channel. HBC also implements secure BAN as, unlike WBAN signals, the HBC signals cannot be snooped on by a wireless attacker

## 2.1 Social Networking during Handshaking using Dynamic HBC between Smartwatches

In many cultures, during a social gathering, *social introduction* is performed by *shaking hands*. In the present day, having a digital-self is very common (Facebook, LinkedIn, Twitter), as is having a smart watch or similar smart wearable devices. Social networking (e.g. Facebook friend request, LinkedIn request) during handshaking could be enabled using HBC (Figure 1a). The information exchange is enabled by using the dynamically formed secure, low-loss conducting path between smart-watches on the wrists of the two handshaking parties. The same dynamic HBC technique could connect any wearable device on person 1 with any wearable device on person 2. Connecting the smartwatches allows for a reliable, short-hop (<1ft), and low-loss connection from one person's wrist to other person's wrist.

Such inter-body information exchange would be applicable in variety of uses cases. For example, in a party, one might want to exchange Facebook friend requests. By turning on a software switch in the smartwatch, a person could allow FB friend request exchange with every person he/she shakes hand with in the party. He/she can then choose to accept or reject the requests individually, at their convenience, on the smart watch or on a computer (synced by the smart watch), by going through the list of received requests. Similarly, in a more professional setting, such as a meeting or a conference, LinkedIn contacts or business cards could be exchanged using dynamic HBC between smartwatches. The amount of information shared could be controlled using an application on the smartwatch.

Unlike WBAN based social networking, where one's information is receivable by everyone around, HBC based social networking allows information exchange only between parties who shake hands. This promises higher probability of the exchanged information being relevant to the parties involved.

## 2.2 Secure Authentication

Modern security terminals often use fingerprints as a method for authenticating a human being. Along with fingerprints, (1) advanced security keys worn on the body (e.g. in a smartwatch, Figure 1b), or (2) person specific unique biomarker could be used for an added layer of security during authentication using touch. Such keys will be unique to the person wearing them and can only be transmitted by touch using dynamic HBC.

## 2.3 ULP Body Area Network

The previous two subsections described the use of dynamic HBC during human-human or human-machine interaction (inter-body). HBC is also a very promising candidate for connectivity between multiple devices on the same person's body (intra-body). WBAN is typically used for connected devices on one's body. However, since this required up-conversion and down-conversion of information to Radio Frequency (RF) WBAN devices tend to be energy-inefficient. Moreover, the propagation loss of RF signal through human body is orders of magnitude higher (Figure 1c) in the RF frequency band (GHz) compared to the baseband signal (MHz). HBC provides a low-loss channel that can pass baseband signal without any up and down conversion, making the circuits ultra-low power (ULP) and the received signal to noise ratio (SNR) high. If the interference (described in details in Section 3.2) can be handled, HBC promises order(s) of magnitude lower power (ULP) connectivity within the human body compared to WBAN.

## 2.4 Secure Body Area Network

An important consideration for the connectivity among devices on a human body is the security of the same. Since this human body network will have a significant amount of personal data, it's utmost important to ensure this data cannot be hacked. In WBAN, this is ensured by encrypting the communicated data. However, the wireless signals can be easily snooped on from a distance (Figure 1c). On the other hand, HBC signals cannot be snooped on, unless the person is physically touched. The security of the WBAN and HBC communication can be written as

$P_{hack,WBAN} = EQ$

$P_{hack,HBC} = EQ \times P_{touch}$

Where, $P_{hack,WBAN}$ and $P_{hack,HBC}$ is the probability of hacking a WBAN and HBC signal, $EQ$ is the encryption quality of the encryption used, and $P_{touch}$ is the probability that the hacker can touch a person for the time required to snoop a HBC signal. Assuming iso-encryption, since, $P_{touch} \ll 1$, it can be seen that the probability of hacking a HBC BAN is significantly lower than the probability of hacking a WBAN. It follows that HBC based body area network are significantly more secure than wireless body area network.

$\frac{P_{hack,HBC}}{P_{hack,WBAN}} = P_{touch}$

$P_{touch} \ll 1 \quad \rightarrow \quad \mathbf{P_{hack,HBC} \ll P_{hack,WBAN}}$

## 3. INTERFERENCE-ROBUST ULP HBC
### 3.1 Channel Models
Active research is under progress on developing robust channel models to represent measured HBC channel characteristics. The distributed nature of the HBC network was modeled in [11]. Two prominent channel models have been proposed, namely capacitive [3],[11] and wave propagation [12] model. In [7], the authors attempts to develop a unified channel model to match measured channel characteristics at multiple frequencies and multiple distances. The goal of this work is to show the efficacy of a resettable integrator based NRZ receiver for interference-robust HBC. In this work we use a simple RC channel model as in [3],[11].

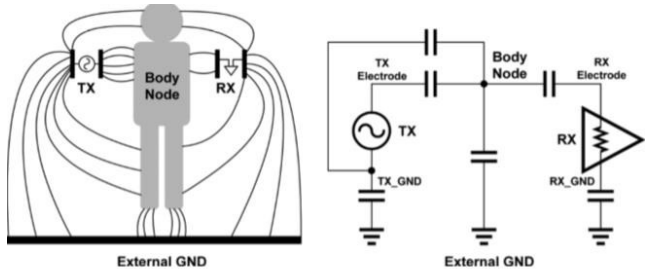

Figure 2: Capacitive Human Body Communication (HBC) and it's effective equivalent electrical circuit **[11]**

### 3.2 Interference on Human Body

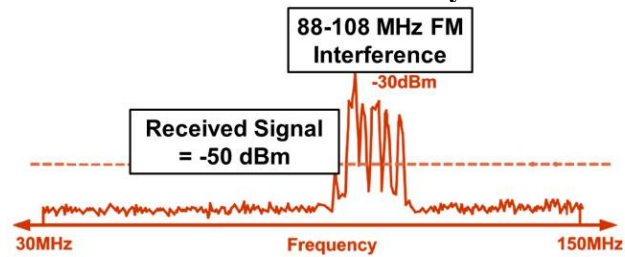

Figure 3: Measured FM interference that affects HBC

One of the biggest challenges in the realization of HBC is that human body acts like an antenna for certain frequencies of interest. This means a ~6ft tall human body will pick up $f_{int} = \frac{c}{2l} = 80\ MHz$, i.e. the human body is susceptible to strong interference for an electromagnetic (EM) signal whose frequency is determined by the wavelength equal to twice of the human height. Similarly, a grounded human body will be susceptible to $f_{int} = \frac{c}{4l} = 40\ MHz$ EM signal as interference. In reality, the human body acts like a lossy conductor leading to a broadband resonance peaking [13]. Hence, the human body acts like an antenna in the 40–400 MHz frequency range.

Incidentally, the FM radio frequency band (88–108 MHz) falls right inside this frequency band. Figure 3 shows an measurement of an example interference spectrum experienced by the body [9], without the Cordless and Walkie-Talkie interference that were intentionally introduced in the original measurement. The FM signals are omnipresent due to its ubiquitous nature. In [9], the authors proposed Adaptive Frequency Hopping (AFH) to avoid this interference issue. The idea is to only transmit and receive in the frequency bands which are not corrupted by the interference. In [10], fixed, two-band narrowband signaling is used. However, both requires modulation and demodulation of multiple narrowband signals on multiple carriers, which tends to be significantly energy-inefficient.

Since human body acts like a conductor, simple Non Return Zero (NRZ) signaling allows the simplest and power-efficient way to communicate data through the body. However, till date the FM interference, as described above has been the biggest bottleneck for such signaling. In this paper, we present an integrating DDR receiver technique along with NRZ signaling that acts as an adaptive notch filter to cut-out the undesired interference. The frequency of the notch can be adjusted simply by choosing the bit rate appropriately.

The following section will develop the theory of integrating DDR receiver and highlight its characteristics as a notch filter.

### 3.3 Signaling
Previous work has proposed signaling techniques like on-off keying (OOK), direct sequence spread spectrum (DSSS), frequency-shift keying (FSK), adaptive frequency hopping (AFH), fixed multi-carrier narrowband signaling etc. All of the aforementioned signaling techniques use narrowband modulation of the signal onto a higher-frequency carrier. In this work we use simple, low-power NRZ signaling. Typically, strong interference has been the bottleneck to such implementations. We propose to suppress the interference using an integrating DDR receiver, as described in details in the following sections.

### 3.4 Theory: Interference Suppression using Integrating DDR Receiver as Notch Filter
Let us consider a NRZ signal with a CW (continuous-wave) interference for simplicity of analysis. The time-domain waveform is shown in Figure 4a. It can be considered as the superposition of the NRZ signal (Figure 4a) and a CW interference signal (Figure 4c). Let us consider that there is an arbitrary phase difference ($\varphi$) between the NRZ and the interference. Often, the interference strength ($A_{intf}$) is significantly larger than the signal amplitude ($A_{sig}$), leading to a closed eye-diagram, making it impossible to sample accurately. Instead, in this work, we propose first integrating the received signal + interference for the bit-period ($T_b$) and then sample. The integration clock, shown in Figure 4d, is a 50% duty cycle clock running at half the bit frequency (i.e. $T_{clk} = 2T_b$). Hence, every other symbol can be integrated using

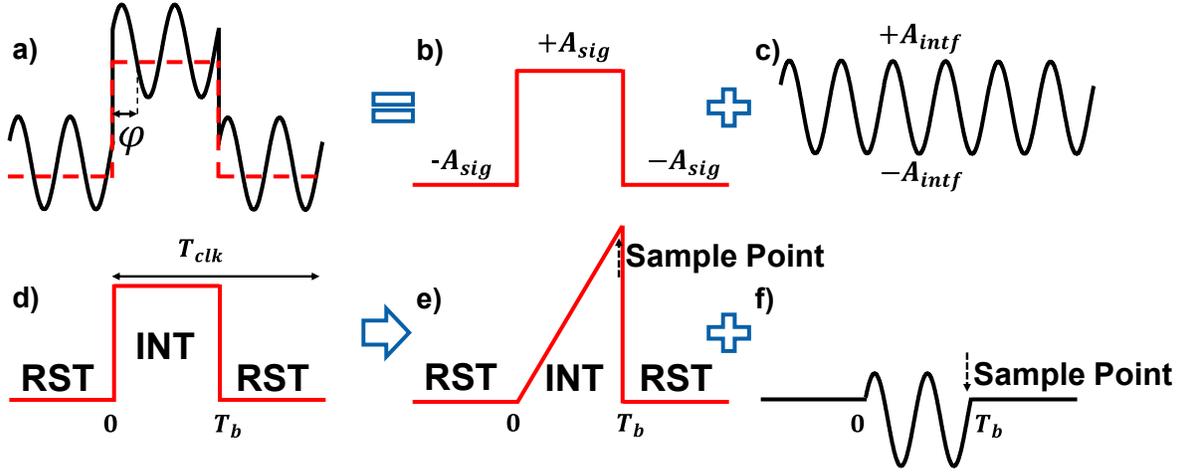

Figure 4: Working principle of Integrating DDR receiver as a notch filter for interference suppression a) NRZ signal + CW interference can be decomposed as a superposition of b) and c). d) shows the integration clock. Each integration period is followed by a reset period, leading to the need for two parallel receiver paths, i.e. dual-data rate (DDR) receiver. $T_{clk} = 2T_b$, e) Integrated NRZ signal f) Integrated interference is 0 for $T_b = nT_i$ (shown for $T_b = 2T_i$)

one phase of the clock. To integrate every symbol, we need two phases of the clock, leading to a dual data rate (DDR) receiver, as shown in Figure 7. The system can be analyzed by writing the received signal ($S_{RX}$) as a linear superposition of desired NRZ signal ($S_{sig}$) and the undesired interference ($S_{intf}$), as follows:

$$S_{RX} = S_{sig} + S_{intf}$$

Now $S_{sig}$ and $S_{intf}$ can be described as:

$$S_{sig}(t) = \pm A_{sig} \quad 0 \le t \le T_b$$
$$S_{intf}(t) = A_{intf} \sin(\omega_i t + \varphi) \quad \forall\, t$$
$$\omega_i = \frac{2\pi}{T_i} = Interference\ Frequency$$

Hence the integrated component ($IS$) of the signal and interference for the 0 clock phase can be written as: ($K_{int} = Integrator\ Gain$)

$$IS_{sig}(t) = \int_0^t S_{sig} = \begin{cases} \pm K_{int} A_{sig} t, & 0 \le t \le T_b \\ 0, & T_b \le t \le 2T_b \end{cases}$$

$$IS_{intf}(t) = \int_0^t S_{intf}$$
$$= \begin{cases} -K_{int}\dfrac{A_{intf}\cos(\omega_i t + \varphi)}{\omega_i}, & 0 \le t \le T_b \\ 0, & T_b \le t \le 2T_b \end{cases}$$

The sampled signal + interference at the end of the bit period i.e. at $t = T_b$ can be expressed as:

$$IS_{sig}|_{t=T_b} = \pm K_{int} A_{sig} T_b$$

$$IS_{intf}|_{t=T_b} = K_{int}\frac{A_{intf}[\cos(\varphi) - \cos(\omega_i T_b + \varphi)]}{\omega_i}$$
$$= K_{int}\frac{A_{intf}\left[\cos(\varphi) - \cos\left(2\pi\dfrac{T_b}{T_i} + \varphi\right)\right]}{\omega_i}$$
$$= 0, \quad \forall\, T_b = nT_i;\ n = positive\ integer$$

It should be noted, if $T_b = nT_i$, for any arbitrary $\varphi$, the contribution of the interference to the integrated and sampled signal will be 0. In other words, by choosing the bit period equal of the NRZ signal as an integer multiple of the period of the interfering signal, the contribution of the interference can be nullified. This forms the fundamental basis for using the integrating DDR receiver as a notch filter to suppress the FM interference in HBC. Figure 4 shows the case where $T_b = 2T_i$. To visualize this better, the integrated interference ($IS_{intf}$) is plotted in Figure 5, for $0 \le t \le T_b$ and varying $\varphi$. The function value varies depending on $\varphi$, for all $t$, except $T_b = nT_i$, where it is 0. The contribution from the interference term given by $IS_{intf}|_{t=T_b}$ and is plotted in Figure 6, showing the relative interference rejection (in dBr) that can be achieved using this technique with varying bit rate ($T_b$), with respect to interference frequency ($1/T_i$). It can be seen that the proposed Integrating DDR Receiver provides >20 dB rejection over the FM band, for a 100Mbps NRZ signal.

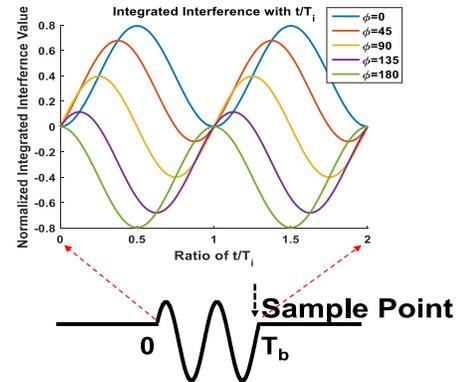

Figure 5: Integrated interference ($IS_{intf}$) as a function of $\dfrac{t}{T_i}$

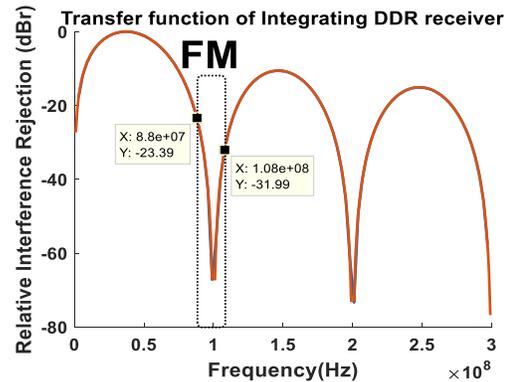

Figure 6: Interference suppression as a function of interferer frequency

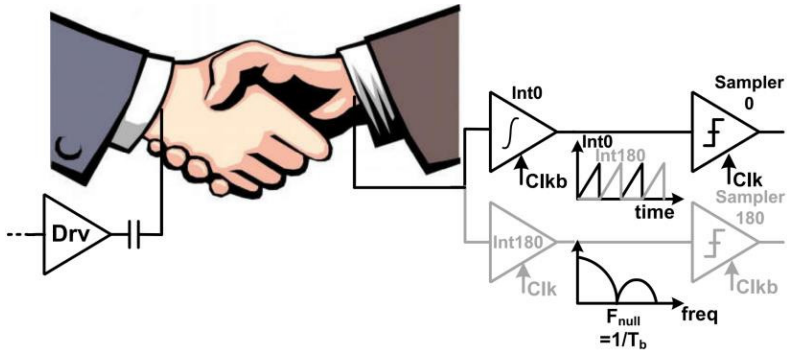

Figure 7: Transceiver diagram for Interference robust HBC. It uses a dual data rate (DDR) receiver with a resettable integrator followed by a sampler

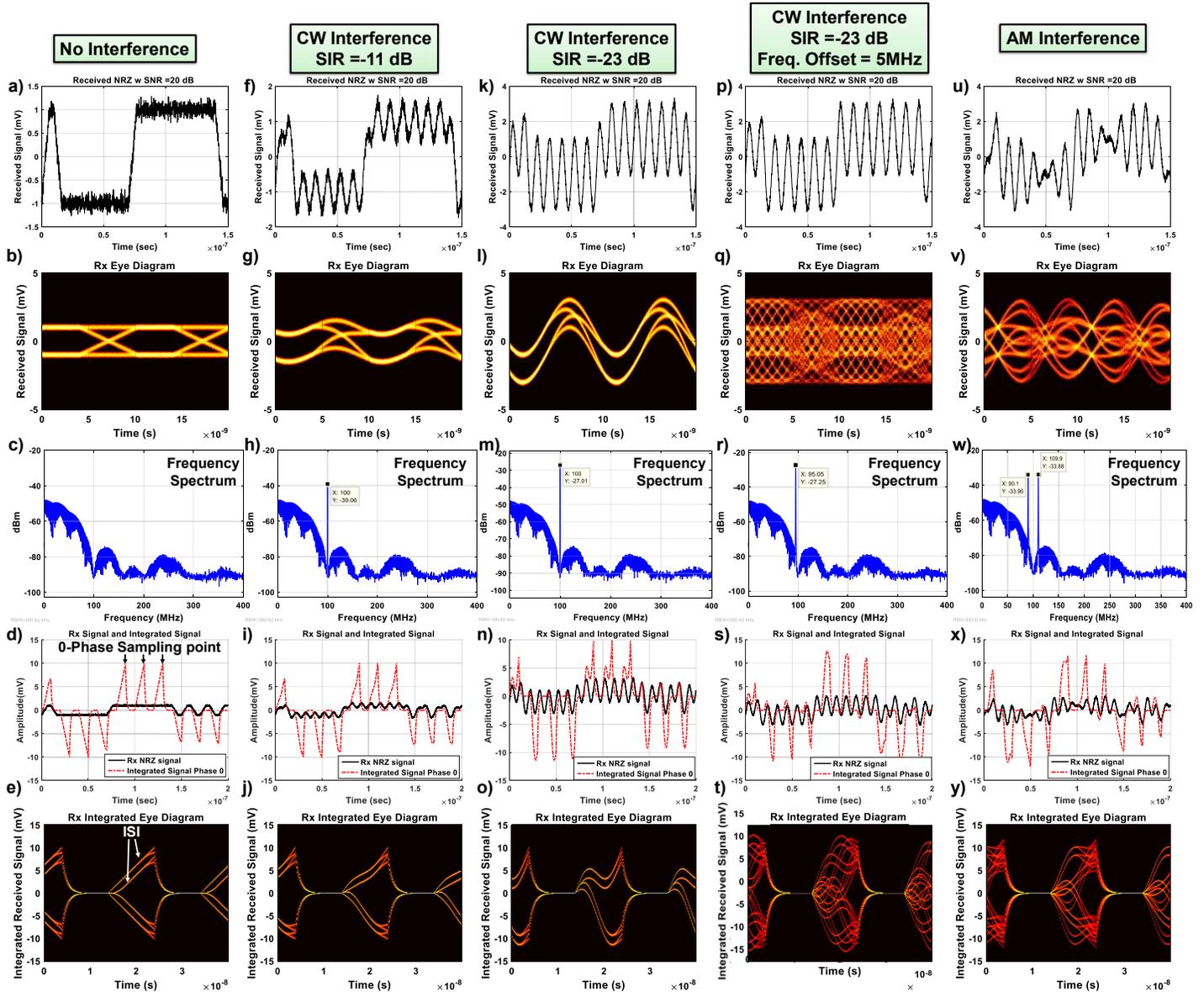

Figure 8: Results for HBC signaling for five different interference cases. NRZ signal with no interference a) Time-domain Rx signal b) Eye-diagram of Rx signal c) Frequency spectrum of Rx signal d) Integrated Rx signal using 0-Phase clock superimposed with Rx signal. The sampling point at the end of the integration period is also highlighted e) Integrated Rx signal showing significant more eye opening. f)-j) Similar plots as a)-e) for NRZ signal plus -39 dBm CW interference (SIR=-11 dB). Integrated eye is significantly less affected compared to the NRZ eye. k)-o) Similar plots as a)-e) for NRZ signal plus -17 dBm CW interference (SIR=-23 dB). Such strong interference results in closing the NRZ eye. However, the integrated eye is still open. This highlights effectiveness of the proposed strong interference. p)-t) Similar plots as a)-e) for NRZ plus -17 dBm CW interference, that is also offset from bit period by 5% (95 MHz CW interference, 100Mbps NRZ data). The NRZ eye is completely closed, but the integrated eye is open. This is due the >20 dB suppression within the 88-108 MHz frequency band provided by the notch. u)-y) Similar plots as a)-e) for NRZ plus AM interference. Similar to previous cases, the integrated eye-diagram performs significantly better than the NRZ eye-diagram.

The above analysis highlights the efficacy of a resettable integrator as a notch filter for a CW interferer. For an AM or FM interference one can arrive at a similar conclusion following a similar but more will not be sampled at $t = T_b$, but at $t = T_b - \delta$, ($\delta$=inverter delay, small). This means that the contribution from interference is very close to 0, but not exactly 0.

## 3.5 Transceiver

The transceiver diagram is shown in Figure 7. A voltage mode driver inside person 1's smartwatch capacitively couples non-return zero (NRZ) signal onto his body. The received signal is picked up by the receiver on person 2's wrist. An integrating receiver is used for interference suppression. Each path integrates for one bit-period ($T_b$) and then resets for next bit period. It follows that two paths (corresponding to 0 and 180 phase of the clock) is needed, i.e. a dual data rate (DDR) receiver is used. The integrated signal is sampled at the falling edge of the integration clock using a strong-arm latch based sampler. Such as receiver suppresses any spurious signal at $f_{null} = 1/T_b$, as derived in the previous section.

## 4. RESULTS

Results for the proposed signaling technique with 100 Mbps NRZ data is shown in Figure 8, for five different interference conditions. Figure 8a to Figure 8e corresponds to a case with NRZ signal with signal to noise ratio (SNR) of 20 dB and no interference present. Figure 8a) shows the time domain noisy received signal, b) shows the eye-diagram of received NRZ signal, c) shows the frequency spectrum of received signal, d) shows the integrated version of the received signal using 0-Phase clock, superimposed with the received signal. The sampling point at the end of the integration period is also highlighted. Figure 8e) shows the eye-diagram of the integrated received signal, showing a larger eye-opening compared to the NRZ eye (Figure 8c). Figure 8f to Figure 8j shows similar plots as Figure 8a-e for NRZ signal plus -39 dBm CW interference (SIR=-11 dB). It should be noted that the integrated eye is significantly less affected by the interference, compared to the NRZ eye. Figure 8k-o, shows similar plots as Figure 8a-e for NRZ signal plus -17 dBm CW interference (SIR=-23 dB). Such strong interference results in closing the NRZ eye. However, the integrated eye is still open. This highlights effectiveness of the proposed strong interference. Figure 8p-t, shows similar plots as Figure 8a-e for NRZ signal plus -17 dBm CW interference, that is also offset from bit period by 5% (95 MHz CW interference, 100Mbps NRZ data). The NRZ eye is completely closed, but the integrated eye is open. This is due the >20 dB suppression within the 88-108 MHz frequency band provided by the notch (Figure 6). Figure 8u-y shows similar plots as Figure 8a-e for NRZ plus AM interference. Similar to previous cases, the integrated eye-diagram performs significantly better than the NRZ eye-diagram. A clock and data recover circuit enable sampling at the maximum opening of the integrated eye-diagram.

## 5. CONCLUSIONS

With the increase of wearable devices, it's becoming utmost important to connect them securely in an energy efficient manner. Human Body Communication (HBC), that uses human body as the communication medium has emerged as a strong contender to wireless body area network (WBAN), due to its attractive features like low-loss, broad bandwidth and inherent security. One of the biggest bottlenecks to successful implementation of HBC is the human body antenna effect that picks up omnipresent FM signals as interference. The interference problem has till date limited HBC implementation to mostly narrowband techniques. In this work, we propose an integrating DDR receiver technique that uses the notch of the transfer function to provide strong suppression of the interference. Interference in the FM band can be suppressed by >20 dB and successful signaling is demonstrated with SIR of -23 dB. The notch frequency can be easily adapted by changing the bit-rate. The proposed signaling technique is supported by both theoretical derivation and system analysis results. Along with the interference-robust HBC signaling technique, we have highlighted promising applications, such as social networking during handshaking, secure authentication, and ULP, secure intra-body BAN using HBC.